\documentclass[aps,prl,twocolumn,
               notitlepage,mathrsfs,
               amssymb,amsmath,amsfonts,
               superscriptaddress,
               longbibliography,
               nofootinbib,floatfix]{revtex4-2}
              
\usepackage[normalem]{ulem}
\usepackage{graphicx}
\usepackage[linktocpage,breaklinks]{hyperref}
\usepackage[capitalize]{cleveref}
\usepackage[usenames,dvipsnames]{xcolor}
\hypersetup{colorlinks=true,
            citecolor=NavyBlue,
            linkcolor=NavyBlue,
            urlcolor=NavyBlue}
\usepackage{multirow,array}
\DeclareMathAlphabet{\pazocal}{OMS}{zplm}{m}{n}
\usepackage{journals}

\newcommand{\dd}{{\rm d}}

\begin{document}
\title{Tidal Resonance in Binary Neutron Star Inspirals: A High-Precision Study in Numerical Relativity}

\date{\today}

\author{Hao-Jui Kuan}
\email{hao-jui.kuan@aei.mpg.de}
\affiliation{Max Planck Institute for Gravitational Physics (Albert Einstein Institute), 14476 Potsdam, Germany}

\author{Kenta Kiuchi}
\affiliation{Max Planck Institute for Gravitational Physics (Albert Einstein Institute), 14476 Potsdam, Germany}
\affiliation{Center of Gravitational Physics and Quantum Information, Yukawa Institute for Theoretical Physics, Kyoto University, Kyoto, 606-8502, Japan}

\author{Masaru Shibata}
\affiliation{Max Planck Institute for Gravitational Physics (Albert Einstein Institute), 14476 Potsdam, Germany}
\affiliation{Center of Gravitational Physics and Quantum Information, Yukawa Institute for Theoretical Physics, Kyoto University, Kyoto, 606-8502, Japan} 

\begin{abstract}
We investigate the tidal resonance of the fundamental ($f$-)mode in spinning neutron stars, robustly tracing the onset of the excitation to its saturation, using numerical relativity for the first time.
We performed long-term ($\approx15$~orbits) fully relativistic simulations of a merger of two highly and retrogradely spinning neutron stars. 
The resonance window of the $f$-mode is extended by self-interaction, and the nonlinear resonance continues up to the final plunging phase.
We observe that the quasi-circular orbit is maintained throughout since the dissipation of orbit motion due to the resonance is coherent with that due to gravitational waves.
The $f$-mode resonance causes a variation in the stellar spin of $\gtrsim6.3\%$ in the linear regime and much more as $\sim33\%$ during the later nonlinear regime. 
At the merger, a phase shift of $\lesssim40$~radians is rendered in the gravitational waveform as a consequence of the angular momentum and energy transfers into the neutron star oscillations. 
\end{abstract}
\maketitle

\textbf{Introduction}---
Binary neutron star (BNS) mergers are among the most important systems to learn about the equation of state (EOS) of nuclear matter through measuring and analyzing the delicate imprint of tidal effects on gravitational waves (GWs); see, e.g., the reviews \cite{Andersson:2010ufc,GuerraChaves:2019foa,Chatziioannou:2020pqz,Dietrich:2020eud,Suvorov:2024cff} and the references therein.
How a neutron star (NS) reponses to the tidal field sourced by its companion is sensitive to its internal structure that is determined by the EOS \cite{Lattimer:2000nx}.
The response function in general depends on the tidal forcing frequency (about twice the orbital frequency) with the zeroth order term being
\begin{align}
    \tilde{\Lambda}=\frac{16}{13M^5}\left[ (M_1+11M_2)M_1^4\Lambda_1 + (M_2+11M_1)M_2^4\Lambda_2 \right],
\end{align}
which is linked with the Love numbers of the participating NSs~\cite{Flanagan:2007ix,Hinderer:2007mb,Hinderer:2009ca,Binnington:2009bb,Damour:2009vw}.
Here $\Lambda_1$ and $\Lambda_2$ are the tidal deformabilities of the two NSs, while $M_1$ and $M_2$ are their respective (Arnowit-Deser-Misner) masses with the total mass of the binary given by $M=M_1+M_2$.
This tidal quantity has been constrained (though not significantly) by the first BNS event GW170817 \cite{Annala:2017llu,LIGOScientific:2018cki,Most:2018hfd,Chatziioannou:2018vzf,De:2018uhw,Landry:2018prl,Narikawa:2019xng}.
As the binary approaches the merger, the frequency-dependent response becomes enhanced to manifest dynamical tides \cite{Steinhoff:2016rfi,Andersson:2019ahb,Andersson:2019dwg,HegadeKR:2024agt}.
This aspect has been incorporated in waveform modeling by interpreting dynamical tides as the excitation of the fundamental ($f$-) mode of steller oscillations \cite{Andersson:2019ahb,Andersson:2019dwg,Hinderer:2016eia,Steinhoff:2016rfi,Ma:2020rak,Kuan:2022etu,Kuan:2023qxo} or the somehow resummed contributions of higher post-Newtonian (PN) orders \cite{Damour:2009wj,Baiotti:2010xh,Bernuzzi:2014owa,Akcay:2018yyh}.

Developing accurate gravitational waveform models is essential to extract the tidal effects from the detected GWs, which requires one to study dynamical tides in detail \cite{Andersson:2017iav,Pratten:2021pro,Williams:2022vct}.
To this end, a variety of analytic efforts has been devoted 
to advancing the knowledge about dynamically perturbed NSs \cite{Gaertig:2008uz,Gaertig:2010kc,Passamonti:2020fur,Pnigouras:2022zpx,Passamonti:2022yqp,Gupta:2023oyy}, 
to incorporating the high PN effects \cite{Vines:2011ud,Damour:2012yf,HegadeKR:2024slr,Mandal:2023hqa}, 
to improving on the modeling of tidal resonance \cite{Ma:2020rak,Kuan:2022etu,Yu:2022fzw,Kuan:2023qxo,Mandal:2022nty,Mandal:2022ufb,Yu:2024uxt}, and 
to devising an adequate effective-one-body treatment for tidal effects \cite{Bernuzzi:2014owa,Hinderer:2016eia,Steinhoff:2016rfi,Nagar:2018plt,Steinhoff:2021dsn}.
However, analytic attempts have limited power in handling the nonlinear tidal response of late-inspiral BNSs, where numerical relativity (NR) is the unique tool to resolve the dynamics involved \cite{Shibata:2005ss,Baiotti:2011am,Bernuzzi:2012ci,Hotokezaka:2013mm,Hotokezaka:2015xka,Hotokezaka:2016bzh,Dietrich:2017feu,Steinhoff:2021dsn,Gamba:2022mgx}. 

The state-of-the-art waveform models for BNSs include:
(I) Tidally-tuned effective-one-body (EOB) models TEOBResumS \cite{Damour:2009wj,Bernuzzi:2014owa,Nagar:2018zoe,Nagar:2018plt,Gamba:2023mww} and SEOBNRv*T \cite{Hinderer:2016eia,Steinhoff:2016rfi,Steinhoff:2021dsn} (see \cite{Rettegno:2019tzh,Dietrich:2017feu} for some comparison surveys); and
(II) point-particle EOB or phenomenological phase models (IMPheom; \cite{Hannam:2013oca,Pratten:2020ceb}) + close form of tide-related phase shift \cite{Dietrich:2017aum,Kawaguchi:2018gvj,Schmidt:2019wrl,Dietrich:2019kaq,Abac:2023ujg,Colleoni:2023czp,Williams:2024twp,Shterenberg:2024tmo}.
The construction of these models appeals to calibration with NR results in one way or another.
Therefore, their validity could be limited to the parameter space covered by the NR waveforms in the literature.
In this spirit, numerical studies of unexplored BNS parameters are essential to guide future waveform modeling.

In particular, mergers of retrogradely spinning NSs are valuable for accurately studying how the $f$-mode resonance develops to a nonlinear phase and its saturation, as the reduced mode frequency \cite{Doneva:2013zqa,Kruger:2019zuz,Kruger:2020ykw,Kruger:2021zta} can ensure resonance before the plunge.
Some literature, such as \cite{Bernuzzi:2013rza,Dietrich:2016lyp,Foucart:2018lhe,Tsokaros:2019anx,East:2019lbk,Dudi:2021wcf}, has simulated BNSs with anti-aligned spins.
However, the spin parameters considered in these studies only allowed for resonance to occur in the merger phase, when the system becomes essentially one body, thus obscuring the $f$-mode resonance effects.
In this Letter, we aim to provide a detailed analysis of $f$-mode resonance, self-consistently demonstrating its nonlinear excitation and back-reaction to the NS.
For this purpose, we performed high-precision long-term general relativity (GR) simulations (covering the last $\approx15$ orbits), enabled by cutting-
edge numerical techniques, to robustly resolve this relativistic hydrodynamical process for the first time.
The geometrical units are assumed throughout this Letter, unless explicitly stated otherwise.

\textbf{$f$-mode effects}---
Dynamical tides can be effectively modeled as a set of spherical harmonic oscillators in linear order, each representing a star's quasi-normal mode (though see \cite{Pitre:2023xsr}).
When the star is immersed in a tidal field exerted by a companion, these oscillators are enforced, with a mode resonating when the tidal frequency ($\Omega_{\rm tid}$) matches its characteristic frequency \cite{Zahn70,Zahn:1977mi}.
For the quasi-circular inspiral and focusing on the dominant tidal effect, the forcing rate approximates the instantaneous frequency ($\omega_{\rm gw}$) of emitted GWs (i.e., $\Omega_{\rm tid}\simeq\omega_{\rm gw}$).
Among the spectra of stellar oscillations, the $f$-mode couples most strongly to the tidal field, and thus its (perhaps resonant; see, e.g., \cite{Mohanty:2024usv} for discussions) excitation will adjust the binary motion by efficiently tapping off the binary's orbital energy to fuel its kinetic energy \cite{Lai:1993di,Shibata:1993qc,Kokkotas:1995xe,Lai:1997wh,Ho:1998hq}.
As a result, the merger will occur noticeably earlier, and the resulting waveform dephasing is important for data analysis of future GW observatories \cite{Pratten:2019sed,Andersson:2019ahb,Pratten:2021pro,Kuan:2022etu,Williams:2022vct}.

If the amplitude of a mode grows enough, the nonlinear self-interaction will become influential \cite{Yu:2022fzw} (see also \cite{Pitre:2025qdf}).
To encode such an effect in the evolution of $f$-mode's amplitude ($c_f$), we phenomenologically introduce self-interacting terms to the Hamiltonian,
\begin{align}\label{eq:ham}
    \mathcal{H}=\frac{1}{2}\dot c_f^2 + \frac{\omega_f^2 c_f^2}{2} + \frac{\eta}{4} c_f^4,
\end{align}
while leaving out the contributions of $\mathcal{O}(c_f^5)$. 
Here, the overhead dot denotes the time derivative, $\omega_f$ is the $f$-mode's frequency, and $\eta$ is a constant setting the degree of the self-interaction.
The associated evolution equation is written as \cite{Kwon:2024zyg}
\begin{align}
    \ddot c_f + \omega_{f,\rm eff}^2 c_f
    = \mathcal{F}_{\rm tid}=A\cos(\Omega_{\rm tid}t)
\end{align}
with $\omega_{f,\rm eff}^2=\omega_f^2+\eta c_f^2$ and the magnitude of tidal force $A$.
If the $f$-mode gets excited to a nonlinear regime such that $\omega_f^2\gg\eta c_f^2$ no longer holds, the Hamiltonian \eqref{eq:ham} manifests an anharmonic oscillator, with the resonance condition given by $\Delta_{\rm eff}^2=\omega_{f,{\rm eff}}^2-\Omega_{\rm tid}^2\approx0$ \cite{Landau1969}.
This nonlinear tidal effect, combined with another nonlinear influence from spin variation in the star \cite{Fuller:2011jb,Yu:2025ptm}, can, in some cases, lock the mode in resonance with the tidal frequency.

In this Letter, we focus on a representative binary in which the involved NSs spin retrogradely with respect to the rotation axis of the system (referred to as the $z$-axis and the equatorial plane of the orbit is assumed to lie on the $x$-$y$ plane).
The $z$-component of the angular momentum of one of the NSs (say NS 1) is computed via [cf.~Eq.~(59) in \cite{Lam:2022yeg}]
\begin{align}\label{eq:loc_s}
    J_1 = \int q_b[ (q_x-h\bar{u}_x)(y-\bar{y}_1) - (q_y-h\bar{u}_y)(x-\bar{x}_1) ] \, \dd\mathcal{V},
\end{align}
where the volume integral is taken over the interior of the NS.
In the above expression, we define $q_b=\sqrt{\gamma}\rho w$ and $q_{x,y}=hu_{x,y}$ with $\gamma$ the determinant of the spatial metric, $\rho$ the rest mass density, $w$ the Lorentz factor, $u_{x(y)}$ the $x(y)$-component of the covariant four-velocity, and $h$ the specific enthalpy, respectively.
The overhead bar denotes the volume-average value for the quantities; for example, $\bar{u}_x$ is the approximate velocity at the mass center of the NS which is located at $(\bar{x}_1,\bar{y}_1)$. 
In GR, the spin of each member of a binary cannot be gauge-invariantly defined. 
In this sense, the analysis based on definition \eqref{eq:loc_s} should be viewed as an approximate indicator.

The energy ($\Delta M_{\rm res}$) and angular momentum ($\Delta J_{\rm res}$) transferred during a tidal resonance in NS 1 will build up a differential rotation and yield a change in its spin as
\begin{align}
    \chi_1+\Delta\chi_1&=\frac{J_1+\Delta J_{\rm res}}{(M_1+\Delta M_{\rm res})^2} \nonumber \\
    &\approx \frac{J_1}{M_1^2}\left( 1 + \frac{\Delta J_{\rm res}}{J_1} - \frac{2\Delta M_{\rm res}}{M_1}
    \right),
\end{align}
where $\chi_1:=J_1/M_1^2$ is a dimensionless spin.
As indicated by the numerical results below, the binary orbit will remain quasi-circular after a resonance (i.e., no sizeable eccentricity is induced).
The variations in the energy and angular momentum can thus be related through the orbital angular velocity $\Omega_{\rm orb}$ as  (see, e.g., \cite{Lai:1993di,Shibata:2004qz}) 
\begin{align}
    \Delta J_{\rm res} \simeq \frac{\Delta M_{\rm res}}{\Omega_{\rm orb}}=\frac{2\Delta M_{\rm res}}{\omega_f},
\end{align}
which in turn yields the fractional change in spin as 
\begin{align}\label{eq:spinup}
    \frac{\Delta \chi_1}{\chi_1}&=2\frac{\Delta M_{\rm res}}{M_1}\left[(M_1\omega_f)^{-1}\frac{M_1^2}{J_1}-1
    \right]\nonumber\\
    &\simeq -7\% \left( \frac{\Delta M_{\rm res}/M_\odot}{5\times10^{-4}} \right)
    \left[ \left(\frac{0.03}{M_1\omega_f}\right)
    \left(\frac{-0.48}{\chi_1}\right)-1 \right].
\end{align}
Note that for $\chi_1<0$, the negative fractional change indicates a positive value of $\Delta\chi_1$.
In addition to the tidally induced oscillations inside the NS, the tidal force torques the NS in an effect to align the tidal bulge on the NS with the external tidal potential.
Consequently, the NS gains angular momentum in the orbital direction \cite{Ogawaguchi:1996kt,Lai:1997wh}.
However, this effect is subdominant to the spin-up caused by $f$-mode resonance.

The above Newtonian picture can be approximately brought into relativistic tidal response~\cite{Hinderer:2016eia,Steinhoff:2016rfi,Steinhoff:2021dsn,HegadeKR:2024agt}, while it is worth noting that this interpretation is an approximation to the response in reality since the quasi-normal modes are not complete in GR \cite{Detweiler:1973,Friedman:1975,Pitre:2023xsr}.
In addition, the quantities defined above depend on gauge and lose strict meaning in GR. 
That said, we will use them as indicators of what is happening in the GR simulation.

\textbf{Numerical Scheme}---
We adopt the code \texttt{SACRA-MPI} \cite{Yamamoto:2008js,Kiuchi:2017pte,Kiuchi:2019kzt}, which employs the Baumgarte-Shapiro-Shibata-Nakamura-puncture formalism \cite{Baumgarte:1998te,Shibata:1995we,Campanelli:2005dd,Baker:2005vv} with the Z4c constraint propagation prescription \cite{Hilditch:2012fp} to integrate Einstein's equation with an adaptive mesh (2:1) refinement algorithm. 
Implementation details are in the cited articles.
For the simulations presented here, the cell-centered grid is configured with four comoving, concentric finer boxes per NS and six coarser domains encompassing both stacks of finer domains.
All domains use a grid of $(2N,2N,N)$ points in the $(x,y,z)$ direction for an even number $N$, where an equatorial mirror symmetry on the $z = 0$ orbital plane is imposed. We choose $N=118,$ 126, 158, 174, and 190, and the finest domain’s size is set to $14.77$~km with a grid spacing denoted as $\Delta x=14.77\,\mathrm{km}/(N+1/2)$.
The key improvements, implemented in the latest version \cite{Kiuchi:2022ubj},  for obtaining an accurate simulation of $f$-mode resonance (see the Supplemental Material~\cite{SupMat} for the evaluation) include the HLLC Riemann solver \cite{Mignone:2005ft,White:2015omx}, refluxing treatment, and a high-order ($6^{\rm th}$) interpolation at the mesh boundary.

To ensure that the $f$-mode resonance is well resolved, we need to consider an inspiraling BNS for which the resonance window is well-separated from the merger phase. 
For this purpose, one is restricted to study BNSs of sufficiently large spin retrograde to the orbit.
In fact, the $f$-mode may not even reach a resonance before the merger if the spin magnitude is lower than a certain threshold, depending on the masses of the binary and the EOS (e.g., \cite{Mohanty:2024usv}).
We consider an equal-mass binary with two NSs spinning with $\chi=-0.48$, where the negative sign denotes the retrograde direction.
To prepare the quasi-equilibrium states of highly-spinning quasi-circular configurations, we use the public spectral code \texttt{FUKA} \cite{Grandclement:2009ju,Papenfort:2021hod} and follow the method sketched in \cite{Buonanno:2010yk} to remove the residual eccentricity until  $e<10^{-3}$.
The initial state is prepared at $M\Omega_{\rm orb}=0.0129$.
We adopt the piecewise-polytropic approximation of the cold EOS SLy4 \cite{Read:2008iy}, and augment a $\Gamma$-law thermal component to it in the manner detailed in, e.g., \cite{Shibata:2005ss}.
We have chosen the index for the thermal component as $\Gamma_{\rm th}=1.8$.
Varying this value will not significantly impact the resonance phenomena studies here since heating effects essentially play no role before the merger phase \cite{Shibata:2005ss} while the resonance occurs before that phase.

\begin{figure}
    \centering
    \includegraphics[width=\columnwidth]{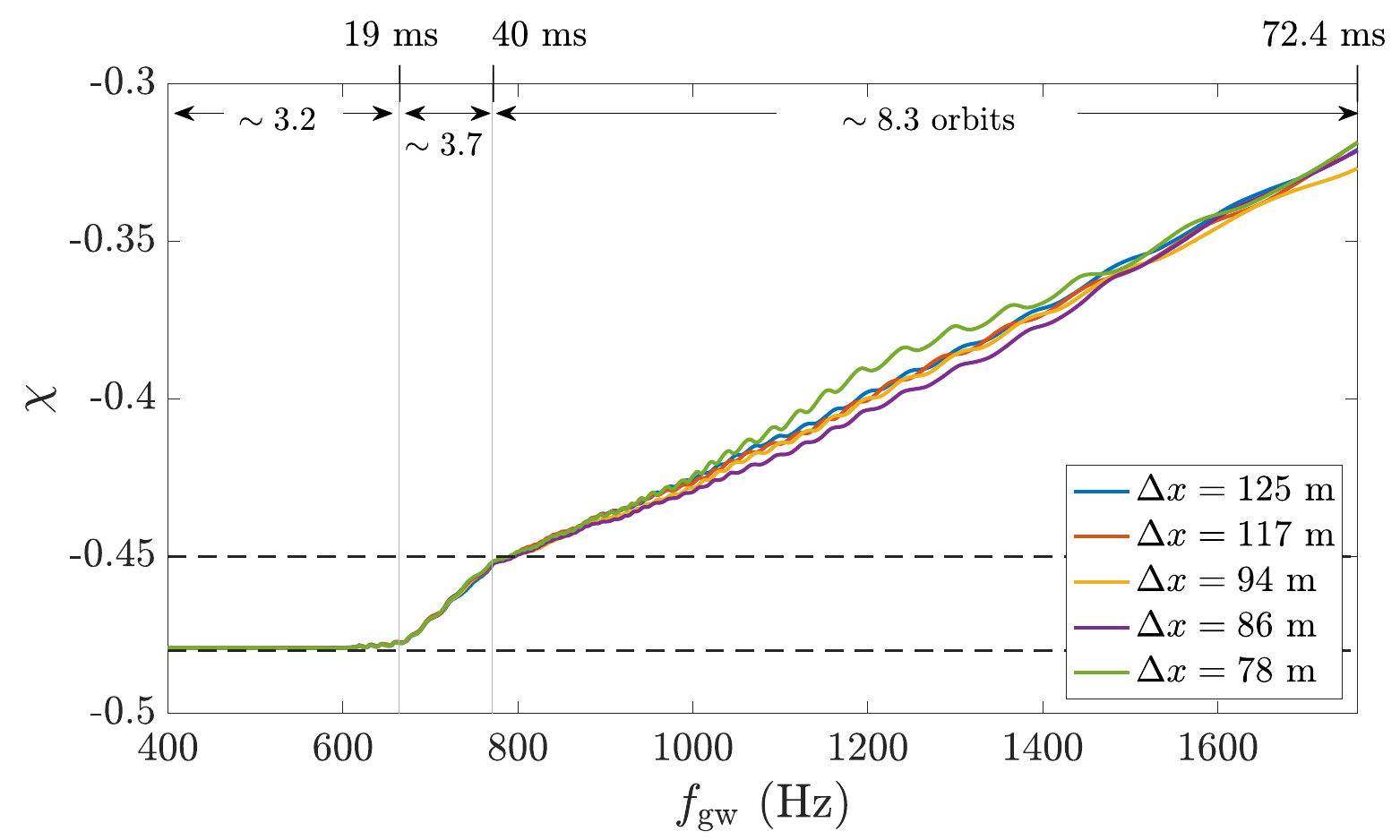}
    \includegraphics[width=\columnwidth]{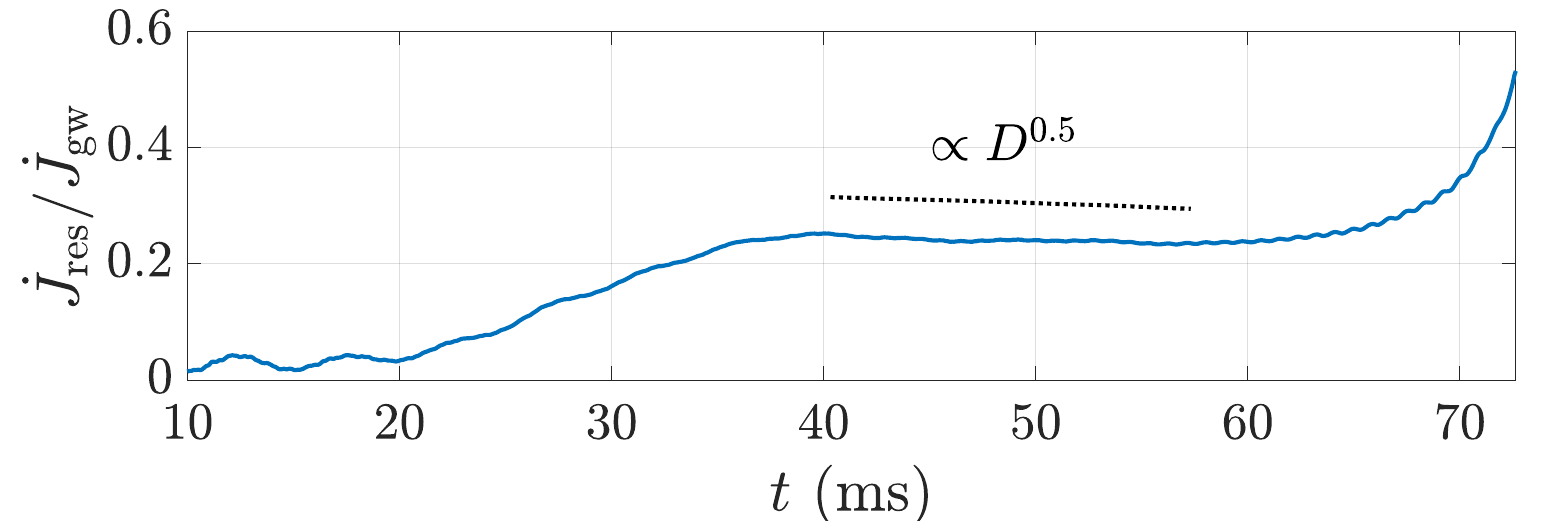}
    \caption{\textit{top}: Local measure of the dimensionless spin [\cref{eq:loc_s}] of an NS in the considered equal-mass BNS as a function of $f_{\rm gw}$ for the grid resolutions adopted. 
    The dashed horizontal lines indicate the dimensionless spin at the onset (lower) and offset (upper) of the linear resonance window, which is independent of the resolutions.
    Time duration and orbits covered before, during, and after this window are also shown.
    \textit{bottom}: The ratio between rates of the angular momentum transfer into an NS via $f$-mode resonance and the angular momentum loss via GWs as a function of time.
    In the bottom panel, we only show the result with the highest resolution adopted.
    }
    \label{fig:s48}
\end{figure}

\textbf{Simulation results}---
The binary evolves $\approx15$ orbits before the merger and the convergence order for the resulted GW phase at the merger is estimated as $p_{\rm conv}\simeq3.851$ ($\lesssim4$; see the supplemental material for the details).
We trace the evolution of the local measure of spin [\cref{eq:loc_s}] for one of the NSs in the top panel of \cref{fig:s48}.
The spin does not vary before the resonance in the first $21$~ms ($\sim3.3$~orbits), implying that the tidal torque is weak even when the orbital separation is $D\lesssim24M$ ($=65$~km).
For the specific spin of $\chi=-0.48$, we estimate the $f$-mode's frequency via the universal relation (5) of \cite{Kruger:2019zuz} as $\omega_f\approx 775$~Hz.
Its resonance begins when $f_{\rm gw}=\omega_{\rm gw}/2\pi\gtrsim670$~Hz, and can be divided into two regimes as indicated by the two slopes of spin-up shown in the evolution of $J$. 
As will be detailed below, these two phases correspond to linear and nonlinear regimes of $f$-mode evolution.
The first epoch lasts for $21$~ms and undergoes $\sim3.7$ orbits, during which the NS's spin increases by $\sim6.3\%$ as indicated by the two horizontal lines in the plot.
According to \cref{eq:spinup}, the star gains the energy of $\Delta M_{\rm res}\sim4.5\times10^{-4}\,M_\odot$ in this phase. 
This energy can also be estimated by the difference in binding energy between the binary at the end of this phase (i.e., at $\Omega_{\rm orb}\simeq390$~Hz) and a quasi-circular, quasi-equilibrium BNS at the same orbital frequency.
In the latter case, orbital decay is due to GW emission, whereas it arises from both GW and tidal excitation in simulation.
The alternative, gauge-invariant estimate gives $\Delta M_{\rm res}\sim8\times10^{-4}\,M_\odot$, agreeing in order of magnitude with the gauge-dependent method.
The absorbed energy drives the $f$-mode oscillation and sustains a degree of differential rotation (see the supplemental material for the profile).

An angular momentum transfer rate accompanies the $f$-mode excitation ($\dot J_{\rm res}$), which is derived from the numerical data and compared with the angular momentum emission rate via GW [$\dot J_{\rm gw}$; Eq.~(2.11) of \cite{Kiuchi:2019kzt}] in the bottom panel of \cref{fig:s48}. 
We see that $\dot J_{\rm res}$ kicks in at $t\simeq20$~ms and grows to $\lesssim 0.25\, \dot J_{\rm gw}$ at $t\simeq40$~ms.
After this point, the ratio between them is broadly `locked' until the merger begins at $ \gtrsim70$~ms.
In particular, the tidal force, scaling as $\propto D^{-3}$ (e.g., \cite{Bildsten:1992my}), continues to efficiently torque the NS through the non-linear evolution of $f$-mode.
The ratio $\dot{J}_{\rm res}/\dot{J}_{\rm gw}$ can then be estimated as $\propto D^{-3}/ D^{-3.5}= D^{0.5}$ during the locking (dotted line).
Summing the contributions of the $f$-mode resonance in two NSs, we see about a $50\%$ increase in the dissipation rate of the binary's angular momentum throughout the nonlinear regime of resonance.
The loss rate of orbital energy is also enhanced by this fraction since the binary remains quasi-circular as we will see below.

After the merger, a black hole promptly forms even though the total mass of the binary is only $M=2.7\,M_\odot$.
This outcome is due to the binary's much less angular momentum compared to an irrotational binary with the same NS masses~\cite{Shibata:2005ss}.
The remnant black hole has a dimensionless spin of $\chi\simeq0.74$, and the irreducible mass of $M_{\rm irr}\lesssim2.41\,M_\odot$.

\begin{figure}
    \centering
    \includegraphics[width=\columnwidth]{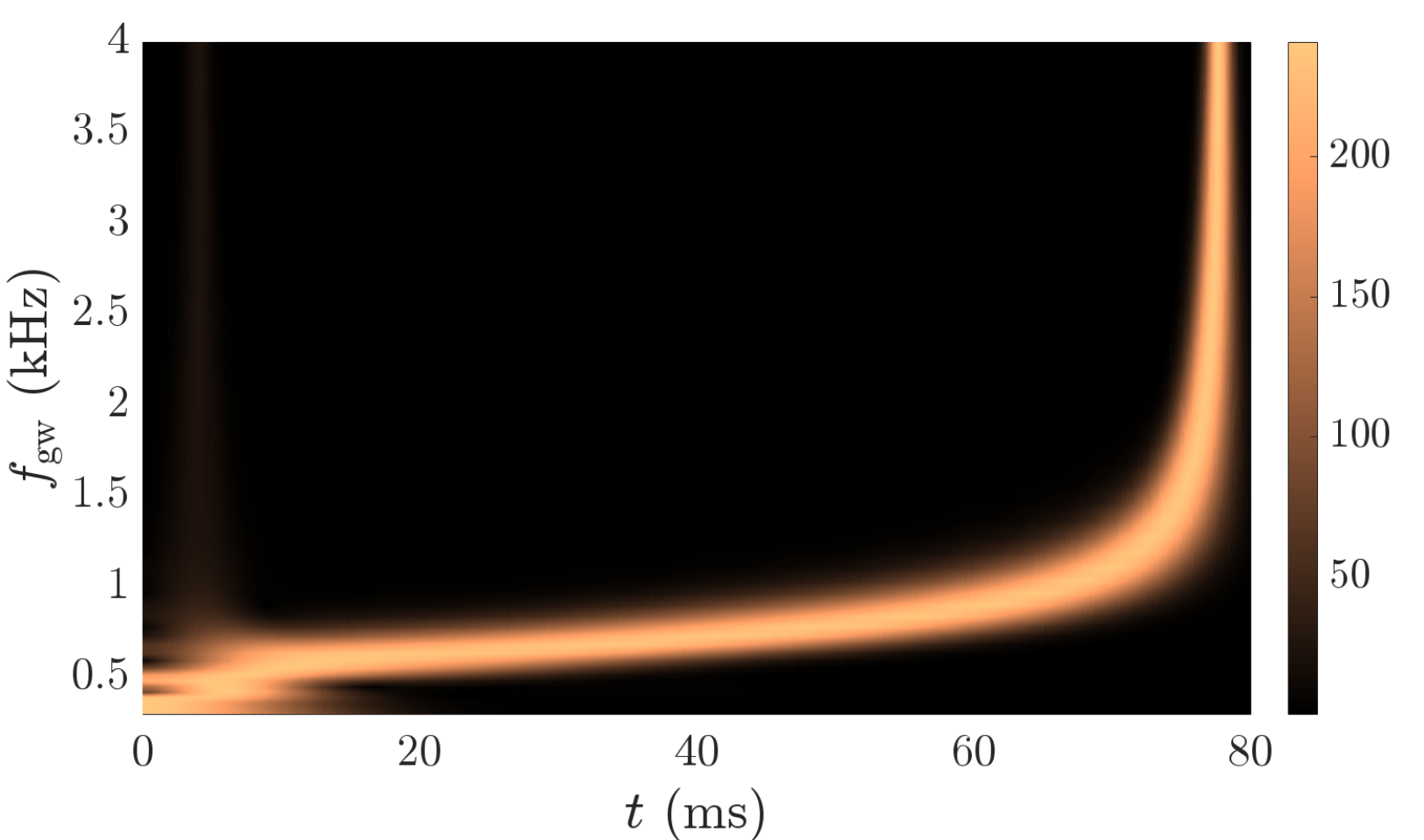}
    \caption{Time-frequency map of numerically computed $\Psi_4$ with the highest resolution adopted. The amplitude scale is arbitrary. In the first 10 ms, low-frequency noise due to initial gauge self-adjustment and junk radiation is observed. After that, the frequency evolution follows a chirp pattern.}
    \label{fig:psi4}
\end{figure}

\textbf{Effect of resonance on GW signal}---
The fixed-frequency method is often adopted to derive waveforms from the numerically computed $\Psi_4$ functions \cite{Reisswig:2010di,Santamaria:2010yb,CalderonBustillo:2022dph}.
This method applies a frequency cutoff to filter out low-frequency noise and spurious contributions below a chosen threshold. 
Recently, \citet{Yu:2024uxt} raised concerns about its accuracy for GWs generated after the onset of $f$-mdoe resonance, suggesting that a non-negligible GW component could result from the excited $f$-mode.
To ascertain that the method is legitimate, i.e., the contribution of $f$-mode emission is negligible, we show in \cref{fig:psi4} the spectrogram of $\Psi_4$ function that is extracted at $r=600\,M_\odot$.
Apart from the blurring in the first 10 ms, which results from the initial data, the signal shows a chirp pattern of the instantaneous frequency up to the merger.
The absent clue of a  $f$-mode-related spectral density validates the use of the fix-frequency method for deriving waveform from $\Psi_4$.
Accordingly, we apply this method to derive waveforms throughout (see \cite{Kiuchi:2017pte,Kiuchi:2019kzt} for the detailed formulae).

\begin{figure}
    \centering
    \includegraphics[width=\columnwidth]{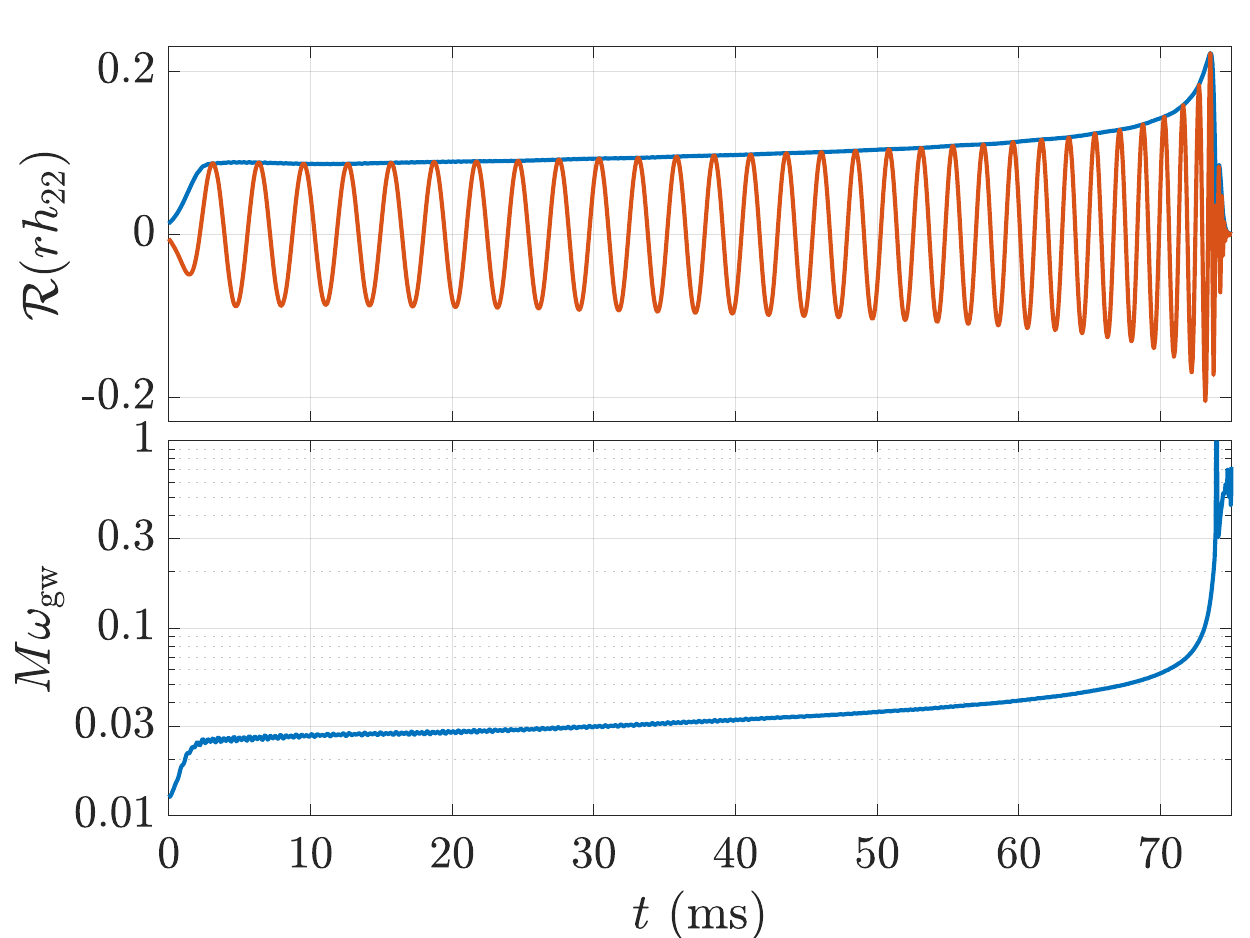}
    \caption{\textit{top}: Waveform covering the last $\approx30$~cycles of inspiral phase (red) and its amplitude over time (blue).
    \textit{bottom}: Evolution of GW angular frequency.
    The shown results are obtained from the highest resolution run.
    }
    \label{fig:strain}
\end{figure}

\begin{figure}
    \centering
    \includegraphics[width=\columnwidth]{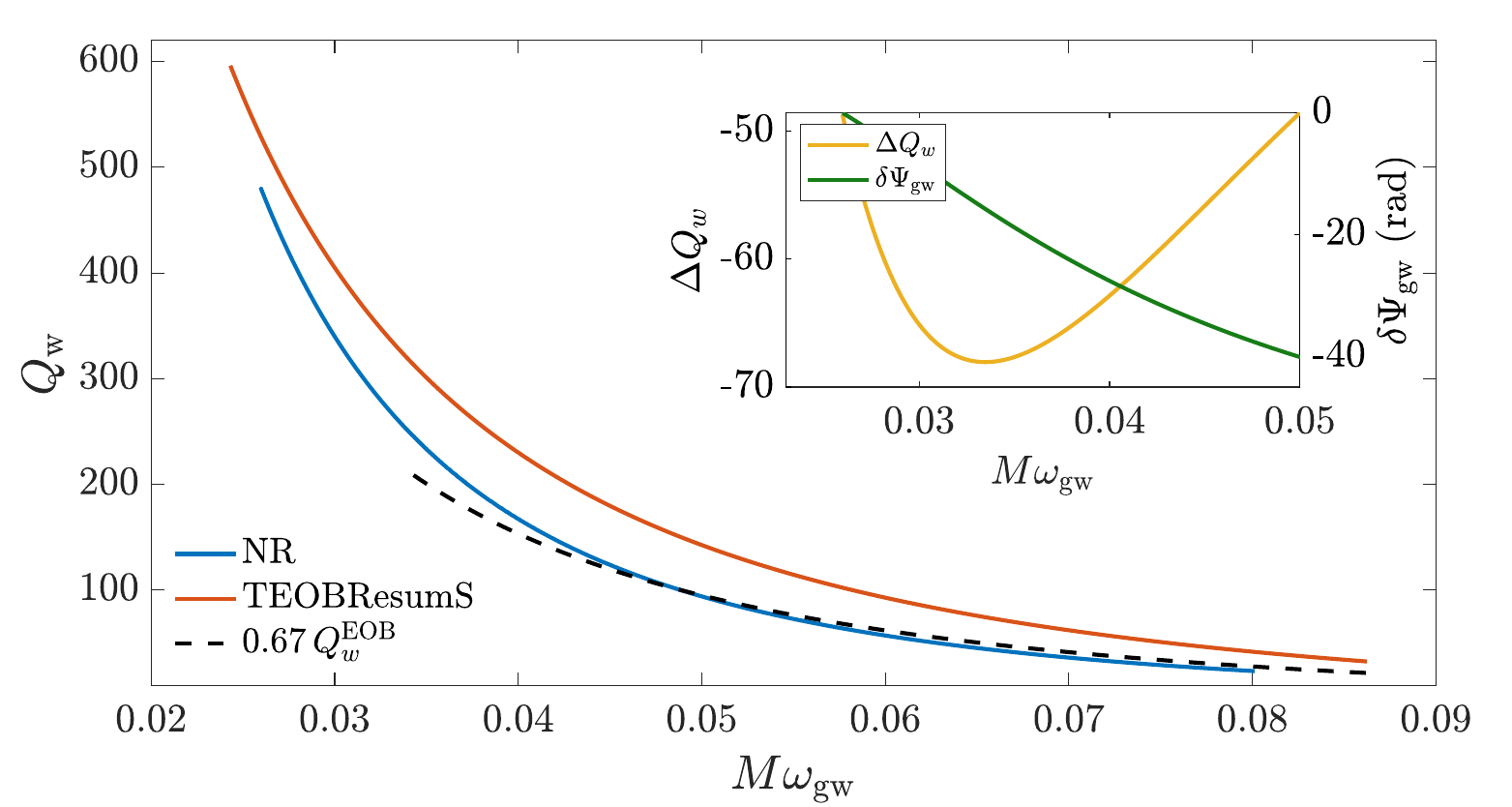}
    \caption{Gauge-invariant and dimensionless measure $Q_w$ of the phase acceleration for the numerical (blue) and the \texttt{TEOBResumS} (red) waveforms.
    The prediction on $Q_w$ modified from the EOB model by the enhanced dissipation due to $f$-mode resonance is overplotted (dashed). 
    The inset window draws the difference between the EOB and NR data (yellow) and the associated dephasing estimated from \cref{eq:dpsi} (green).
    }
    \label{fig:Qw}
\end{figure}

The strain of the $(2,2)$ component of GWs (top) and its time-frequency representation (bottom) are shown in \cref{fig:strain}.
We can see that the $f$-mode resonance induces essentially no eccentricity regardless of the linear or nonlinear regimes. 
The orbit largely remains quasi-circular up to the merger phase as no oscillatory behavior is observed for either the amplitude or the frequency of the waveform. 
The quasi-circularity after the resonance is further evidenced by the acceleration rate of GW's phasing, viz.~$Q_w\equiv \omega_{\rm gw}^2/\dot \omega
_{\rm gw}$ \cite{Baiotti:2010xh,Baiotti:2011am,Bernuzzi:2012ci,Damour:2012ky}.
We diagnose $Q_w$ of the numerical waveform and compare it to the EOB waveform model \texttt{TEOBResumS} in \cref{fig:Qw} while noting that the difference between \texttt{TEOBResumS} and \texttt{SEOBNRv*T} models is much less than the deviation between NR result to them for this BNS configuration.
We use the Love number of a static NS with $M=1.35\,M_\odot$, which is $\Lambda=389.4$, to generate this EOB waveform.
Their overall agreement is poor, and the inset window shows the difference $\Delta Q_w=Q^{\rm NR}_w-Q^{\rm EOB}_w$ (yellow) between the NR result and the EOB data. 

To identify the primary source of deviation, we compare $Q_w$ of this EOB model with an openly available numerical binary black hole waveform (serial number 1500) from the database of the Simulating eXtreme Spacetimes (SXS) collaboration \cite{Mroue:2013xna,Boyle:2019kee} with similar parameters. 
In particular, we generate a \texttt{TEOBResumS} waveform and compare it with the SXS data for an equal-mass binary with both black holes spinning at $\chi=-0.4842$. 
While we do not present these results here, we found that the deviation $\Delta Q_w$ oscillates around zero with a peak value $\simeq4$ all the way up to the merger.
The deviation shown in \cref{fig:Qw} suggests a more pronounced discrepancy when compared to BNS mergers with the given spin parameter. 
We now turn to argue that $f$-mode resonance can result in this deviation.

During the resonance, a portion of the binary's orbital energy is transferred into the NSs.
This energy transfer leads to an increased amount of $\dot\omega_{\rm gw}$, and thus $Q_w$ is lower than the situation where the resonance is absent.
Given the sustained quasi-circularity of the orbit, the $\dot{J}_{\rm res}$-$\dot{J}_{\rm gw}$ ratio (cf.~\cref{fig:s48}) suggests an approximately $50\%$ increase in the energy loss rate of the orbit ($\mathcal{F}$). 
This effectively enhanced $\mathcal{F}$ can be translated to a decrease in $Q_w$ by a factor of $\approx 1.5$ using the energy balance equation $\dot\omega_{\rm gw}=-\mathcal{F}/(\dd E/\dd\omega_{\rm gw})$, where $E$ is the binding energy of the binary.
In \cref{fig:Qw}, we overplot a curve of $0.67 \,Q_w$, which $Q_w^{\rm NR}$ indeed asymptotically approaches\footnote{Although the EOB models \texttt{TEOBResumS} and \texttt{SEOBNRv*T} incorporate the tidal resonance effects to some extent, reproducing the NR results presented here is beyond their current capability.
This limitation arises because they have not been informed by NR waveforms in which the resonance occurs well before the merger phase.}.
Although the local measurement of $\dot{J}_{\rm res}$ is a gauge-dependent quantity, the observed impact on $Q_w$ can be explained by the orbital dissipation enhanced by it.
We also estimate an overall phase shift in the waveform due to the resonance through
\begin{align}\label{eq:dpsi}
    \delta\Psi_{\rm gw}=\int \Delta Q_w \dd(\ln\omega_{\rm gw}),
\end{align}
which is depicted in the inset of \cref{fig:Qw} (green).
The phase shift due to the $f$-mode resonance is enormous and amounts to $\lesssim6.4$ GW cycles (or $\lesssim 3.2$ orbits).

\textbf{Outlook}---
We reported the first GR simulation for $f$-mode resonance that is performed by a new implementation presented in~\cite{Kiuchi:2022ubj}.
Our results underscore the importance of $f$-mode effects for accurate GW data analysis and offer valuable insights into refining current waveform models in the tidal resonance sector.
In particular, we show that the resonance window can be robustly resolved, which reflects an intrinsic feature of tidal interactions in GR. 
Information about its back-reaction to the NS and how it develops from the linear regime and saturates to the nonlinear stage can thus provide hints for the future modeling of other systems, such as eccentric BNS (e.g., \cite{Gold:2011df}) or close encounter resonance.
However, the locking phenomenon between nonlinear resonance and GW dissipation is specific for quasi-circular binaries.
Assuming that the influence of the resonance in the two NSs could be linearly added to render the gross effects, our result can also apply to black hole-NS binaries.

This Letter demonstrates that the numerical scheme has now achieved the precision levels that allow us to explore the dynamical tides with high fidelity. This advancement opens a new path for systematically investigating the $f$-mode resonance.
The Letter also presents an important initiative to complete the NR dataset in the untouched BNS parameter space, which will be key to next-generation waveform modeling of BNS mergers.

\section*{Acknowledgement}
H.-J. K. is indebted to the stimulating discussion with members of the Computational Relativistic Astrophysics division in AEI.
Numerical computations were performed on the Sakura, Raven, and Viper clusters at the Max Planck Computing and Data Facility. This work was in part supported by Grant-in-Aid for Scientific Research (grant No. 23H04900 and 23H01172) of Japanese MEXT/JSPS. 
The numerical simulation of this work was performed on the computational resources of the supercomputer Fugaku provided by RIKEN through the HPCI System Research Project (Project ID: hp230534, hp240532, hp240039, and hp250066). 

\bibliography{references}

\clearpage
\appendix*

\section{Convergence}\label{conv}

\begin{figure}
    \centering \includegraphics[width=\columnwidth]{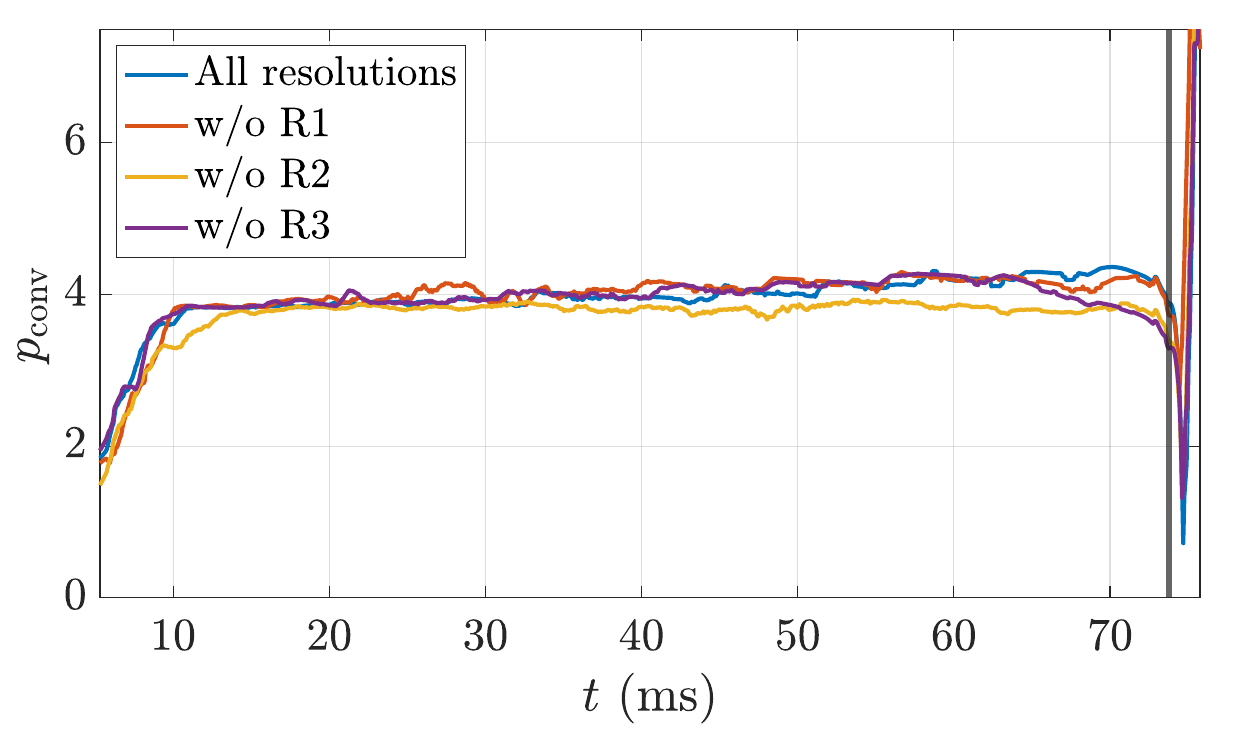}
    \caption{Convergence order for GW phase obtained by simulations of the five resolutions adopted. Estimates are derived from the complete dataset (blue), and also from three subsets of results (discarding one of the lower resolutions at a time; see legend). The merger time of the simulation of the highest resolution R5 is represented by the vertical line.
    }
    \label{fig:pconv}
\end{figure}

We have performed simulations for five resolutions, labeled as R1--5 from the lowest one to the highest.
The finest grid spacing of a given resolution R$_i$ is denoted as $\Delta_{{\rm R}_i}$ with the specific values given by $\Delta_\text{R1--5}\in\{125,117,94,86,78\}$~m.
We use the waveform phasing as the agency to estimate the convergence of numerical results.
Notably, we determine the best fit parameter $p_{\rm conv}$ for the ansatz,
\begin{align}\label{eq:pconv}
    \phi(t; {\rm R}_i)-\phi(t; {\rm R5}) = a(t)\left(\frac{\Delta_{{\rm R}_i}}{\Delta_{{\rm R5}}}\right)^{p_{\rm conv}}\,,
\end{align}
and designate it as the convergence order.
In \cref{fig:pconv}, we depict the evolution of $p_{\rm conv}$ estimated using data from all five resolutions.
To demonstrate the well-definedness of the convergence ansatz in \cref{eq:pconv}, we also provide estimates obtained by excluding one of the lower-resolution runs (R1, R2, or R3) individually.
We observe that the convergence remains approximately fourth order throughout most of the simulation.
However, it slightly decreases to $p_{\rm conv}\alt3.851$ at the merger time of the run at resolution R1.

The estimated $p_{\rm conv}$ meets the expectation based on our numerical implementation.
The decay of orbit is primarily determined by the radiation backreaction, and thus the convergent behavior of waveforms is expected to depend largely on the accuracy of the solver for Einstein's equation.
A 6th-order finite difference treatment prolonged a finer box to a coarser one, and a 4th-order one restricted it in the opposite direction.
For the time integrator, we adopted the 4th-order explicit Runge-Kutta scheme.
A convergence order of 3--4 is thus expected from our numerical setup.
We remark that the convergence order of $p_{\rm conv}$ is also found in the simulations with an independent code \texttt{BAM} \cite{Kuan:2025bzu}.
The hydrodynamics involved inside NSs also play a role in shaping the waveforms through tidal deformation as well as f-mode resonance, though subleading to the contribution of the evolving quadrupole moment of the whole system.

For the hydrodynamics solver, we adopted the 2nd-order HLLC Riemann solver, augmented by a 3rd-order piecewise parabolic method \cite{Colella:1982ee,Shibata:2005gp} to reconstruct hydro-fluxes at cell interfaces.
The convergence order for the matter effects, especially their imprints on the GW signal, should be $\alt 2$ in the absence of strong shock generation.
In order to separately estimate the convergence power of the minor contribution of finite-size effects, we assume an extended expression to \cref{eq:pconv} as
\begin{align}
    \phi(t; {\rm R}_i)-\phi(t; {\rm R5}) = a(t)\left(\frac{\Delta_{{\rm R}_i}}{\Delta_{{\rm R5}}}\right)^{p_{\rm conv}}
    +b(t)\left(\frac{\Delta_{{\rm R}_i}}{\Delta_{{\rm R5}}}\right)^{q_{\rm conv}}\,.
\end{align}
We indeed obtained $p_{\rm conv}\simeq3.8$ and $q_{\rm conv}\simeq1.8$, and found that $a(t)$ exceeds $b(t)$ by several orders of magnitude with $\log a\simeq7.7$ and $\log b\simeq2.2$.
Owing to the predominant contribution of $a(t)$, the overall accuracy of numerical waveforms can be estimated as $p_{\rm conv}$.

The simulation at R5 was done through 20 jobs on the Max Planck Computing and Data Facility's cluster, \textsc{Raven}, with each job occupying 64 nodes. 
Given that each node is equipped with 72 CPU cores, the computational cost of this run approximately totals 2.2 million CPU hours.

\section{Results with the HLLE Riemann Solver}\label{hlle}

\begin{figure}
    \centering
    \includegraphics[width=\columnwidth]{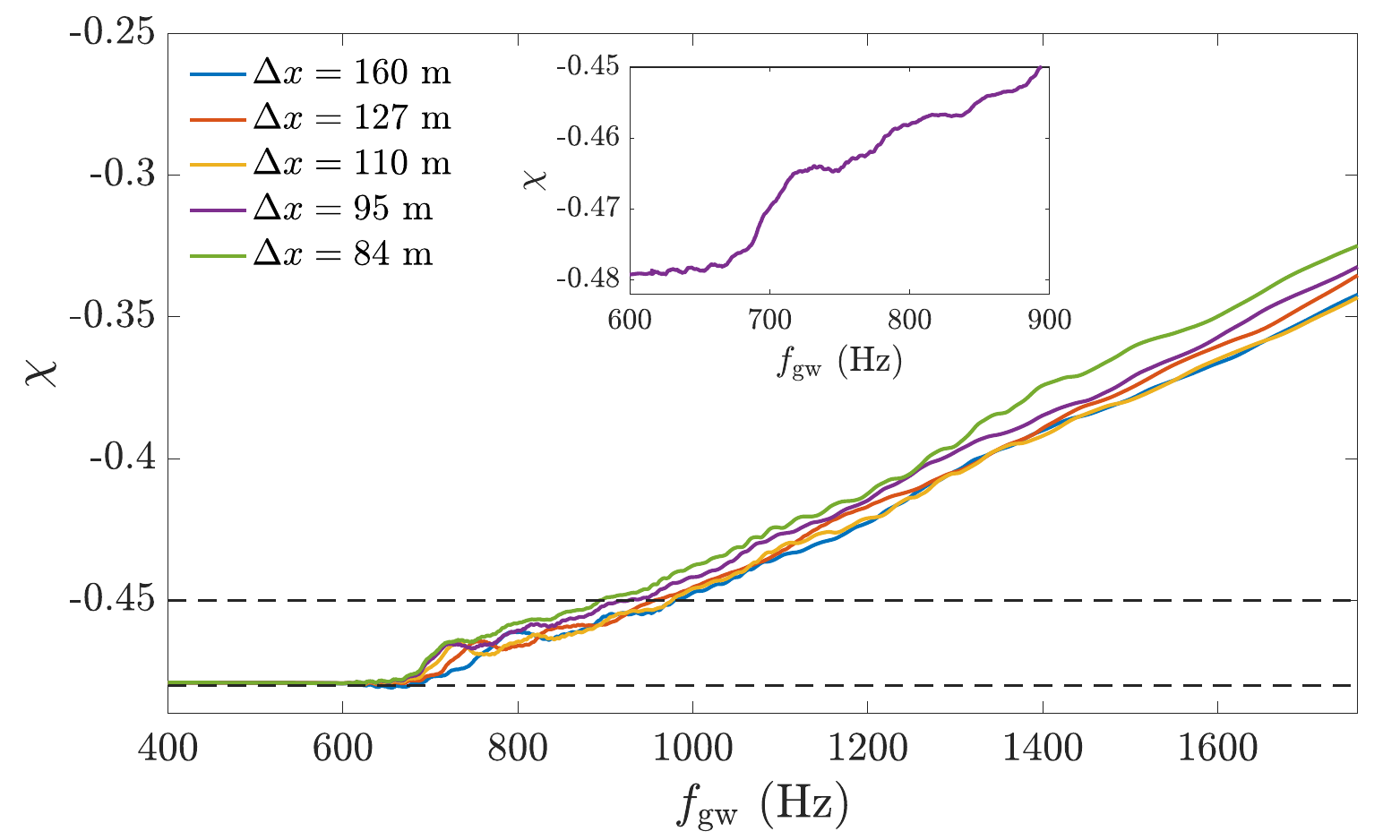}
    \caption{Angular momentum evolutions for one NS with various resolutions (see plot legends). The same initial data as that used in the main text is adopted, while the hydrodynamics is carried out with the HLLE Riemann solver.
    The dashed lines are the same as those depicted in Fig.~1 of the main text for the sake of comparison between the results here and there. The inset window singles out the result with the second-highest resolution adopted.
    }
    \label{fig:hlle}
\end{figure}

Here we provide the simulation results using the HLLE Riemann solver \cite{Einfeldt88}, a popular method in NR for BNS merger studies. 
The grid structure is the same as that described in the main text only that the grid is now vertex-centered.
The resolutions employed here differ from those used earlier. To avoid potential confusion, we do not assign labels to the resolutions in this section; instead, each resolution is represented by its finest grid spacing $\Delta x$.
As illustrated in \cref{fig:hlle}, the angular momentum of one NS remains almost unchanged during the pre-resonance stages, then begins to increase at a frequency that is consistent across the adopted resolutions here and aligns well with the results presented in the main text.
However, the convergence in the growth rate and in the frequency at which the resonance transits from its linear to nonlinear regimes was not achieved.
In addition, this quantity exhibits spurious oscillatory behavior until a resolution as high as $\Delta x=84$~m is adopted for simulation; the inset window emphasizes this behavior even for the simulation using $\Delta x=95$~m.
We also comment on the convergence order estimated at the merger time based on \cref{eq:pconv}, which is found as $p_{\rm conv} \lesssim 1$. 
The much lower convergence order suggests that the HLLE results did not accurately capture the realistic scenario, at least for the resolutions considered here.
The advantage of using the HLLC solver than the HLLE solver for relativistic hydrodynamics is also demonstrated by other numerical test problems in \cite{Lam:2025pmz}.

\begin{figure}
    \centering
    \includegraphics[width=\columnwidth]{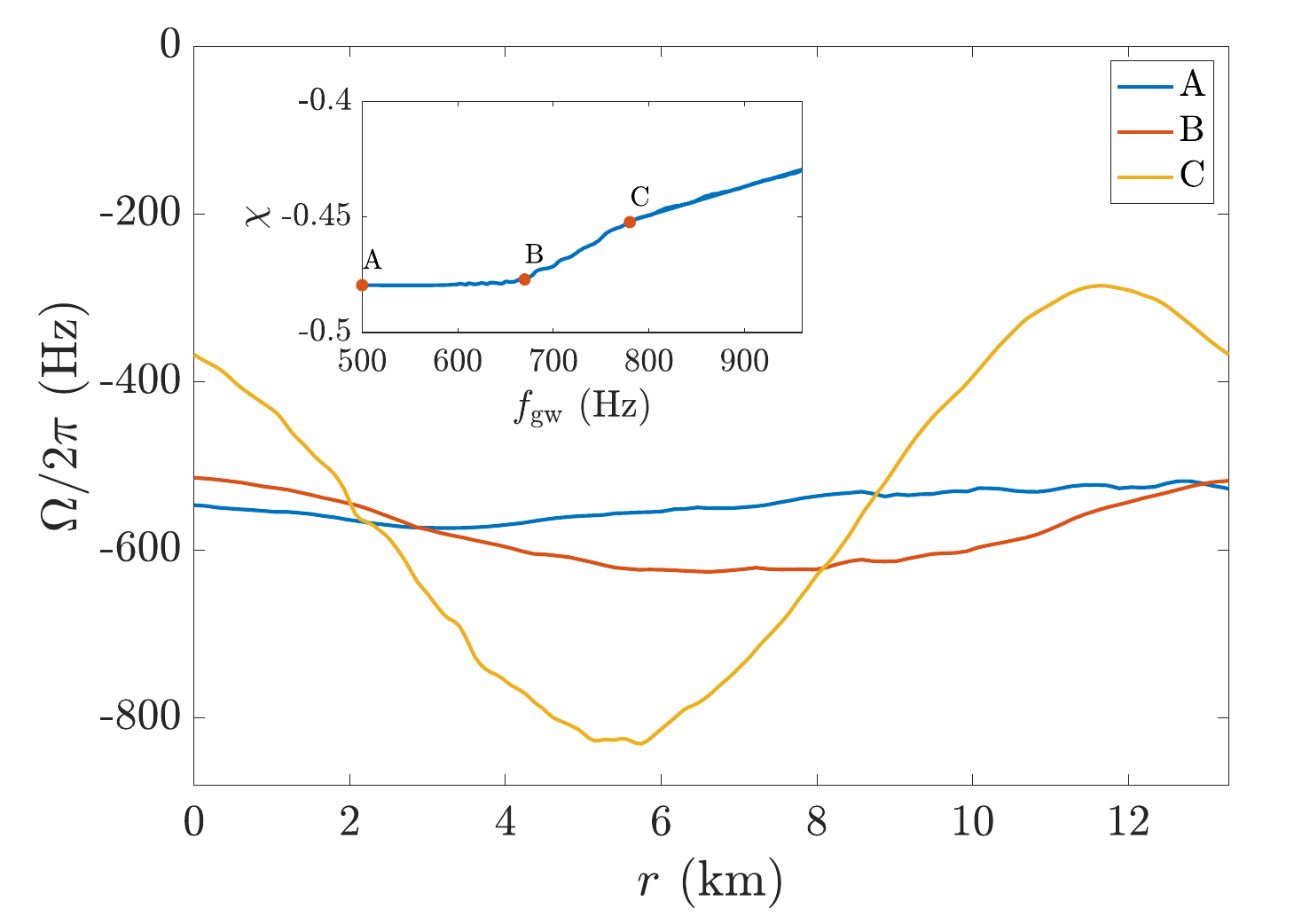}
    \caption{Distribution of stellar rotation rate for one NS in the considered binary at three moments: the initial stage of the simulation ($A$; blue), at the onset of the linear resonance when $f_{\rm gw}\simeq 670$~Hz ($B$; red), and at the end of the linear resonance when $f_{\rm gw}\simeq 780$~Hz ($C$; yellow). The three moments are designated on the evolution of the dimensionless spin $\chi$ in the inset window.
    }
    \label{fig:diff_rot}
\end{figure}

\section{Differential rotation after resonance}\label{diff}

During the $f$-mode resonance in the considered BNS, the retrogradely spinning NSs gain energy $\Delta M_{\rm res}$ and angular momentum $\Delta J_{\rm res}$.
The added angular momentum aligns with the orbital one, thus reducing the magnitude of the NS's angular momentum.
The reader might well be wondering why a spinning-down NS absorbs energy.
In addition to fueling the oscillation generated by the excited $f$-mode, this energy also sustains the NS's differential rotation profile: For a given angular momentum and rest mass, the gravitational mass of a differentially rotating NS exceeds that of a uniformly spinning one.
\cref{fig:diff_rot} shows the angular velocity ($\Omega = u^{\phi}/u^t$) for one of the NSs in the simulation at three specific moments, i.e., the start time of the simulation and the transition times to the linear and the nonlinear resonance regimes (cf. the inset plot). For this, we estimate the angular four-velocity around the NS center (removing the orbital motion residue) via
\begin{align}
    u^{\phi}= \frac{(y-\bar{y}_1)(u^{x}-\bar u^{x})-(x-\bar{x}_1)(u^{y}-\bar u^{y})}
    {(x-\bar{x}_1)^{2}+(y-\bar{y}_1)^{2}},
\end{align}
where we recall that $\bar x(y)$ and $\bar u^{x(y)}$ are, respectively, the volume-averaged values for the position and the velocity of the NS.
In addition, we perform a time averaging on the rotation profile so as to suppress the contribution coming from oscillations.
The averaging is taken over the expected period of excited $f$-mode, i.e., $P\simeq1.3$~ms.

A detailed hydrodynamic analysis is needed to understand the generation of differential rotation and to determine how much portion of $M_{\rm res}$ goes into oscillation and NS interior, respectively.
This issue will be explored in future work.

\end{document}